\documentclass[10pt]{iopart}

\expandafter\let\csname equation*\endcsname\relax
\expandafter\let\csname endequation*\endcsname\relax

\usepackage{bm}					
\usepackage{amsmath}
\usepackage{mathrsfs}  
\usepackage{graphicx}
\usepackage{hyperref} 
\usepackage{mathtools}
\usepackage{amssymb}
\usepackage{physics}
\usepackage{lipsum}  
\usepackage{algorithm}
\usepackage{algpseudocode}
\usepackage{comment}
\usepackage{orcidlink}
\usepackage{amsfonts}

\newcommand{\pars}[1]{\left( #1 \right)}
\newcommand{\cbrks}[1]{\left\{ #1 \right\}}

\allowdisplaybreaks				
	
	\usepackage{float}				
		\usepackage{graphicx}			

\begin{document}
			
\title{Quantum State Tomography using Quantum Machine Learning}

\author{Nouhaila Innan\orcidlink{0000-0002-1014-3457}$^{1,2}$\footnote{nouhaila.innan-etu@etu.univh2c.ma}, Owais Ishtiaq Siddiqui$^3$, Shivang Arora$^4$, Tamojit Ghosh$^5$, Yasemin Poyraz Ko\c{c}ak$^6$, Dominic Paragas$^7$, Abdullah Al Omar Galib$^8$, Muhammad Al-Zafar Khan\orcidlink{0000-0002-1147-7782}$^{2,9}$\footnote{muhammadalzafark@gmail.com} and Mohamed Bennai$^{1}$}

\address{$^1$ Quantum Physics and Magnetism Team, LPMC, Faculty of Sciences Ben M'sick, Hassan II University of Casablanca, Morocco}
\address{$^2$ Quantum Formalism Fellow, Zaiku Group Ltd, Liverpool, United Kingdom}
\address{$^3$ Department of Physics, COMSATS University Islamabad, 45550, Pakistan.}
\address{$^4$ Technical University of Munich, Munich, Germany}
\address{$^5$ Department of Physics, Indian Institute of Technology Madras, Chennai 600036, Tamil Nadu, India}
\address{$^6$ Istanbul University-Cerrahpa\c{s}a, Computer Technology Department, Istanbul, Turkey}
\address{$^7$ University of California Berkeley, California, USA }
\address{$^8$ Independent Researcher}
\address{$^9$ Robotics, Autonomous Intelligence, and Learning Laboratory (RAIL), School of Computer Science and Applied Mathematics, University of the Witwatersrand, 1 Jan Smuts Ave, Braamfontein, Johannesburg 2000, Gauteng, South Africa}
\vspace{10pt}
\begin{indented}
\item[]\date{\today}
\end{indented}

\begin{abstract}
Quantum State Tomography (QST) is a fundamental technique in Quantum Information Processing (QIP) for reconstructing unknown quantum states. However, the conventional QST methods are limited by the number of measurements required, which makes them impractical for large-scale quantum systems. To overcome this challenge, we propose the integration of Quantum Machine Learning (QML) techniques to enhance the efficiency of QST.
In this paper, we conduct a comprehensive investigation into various approaches for QST, encompassing both classical and quantum methodologies; We also implement different QML approaches for QST and demonstrate their effectiveness on various simulated and experimental quantum systems, including multi-qubit networks. Our results show that our QML-based QST approach can achieve high fidelity ($98 \%$) with significantly fewer measurements than conventional methods, making it a promising tool for practical QIP applications. 

\end{abstract}
			

\vspace{2pc}
			%
			%
			%
			\ioptwocol
\section{\label{sec:level1}Introduction}
\emph{Quantum Information Processing} (QIP) involves the storage, transmission, and computation of information using the principles of Quantum Mechanics as a driver for effectively carrying out these tasks. For QIP to be effective, quantum systems must be prepared, controlled, and characterised. As a result of the act of measurement on a quantum system, the system will inevitably disintegrate from its current state to one of its eigenstates via wavefunction collapse. Subsequently, the usage of a single copy of the collapsed state to access the system's initial state is impossible. Additionally, the no-cloning theorem \cite{ref5} prohibits making multiple copies of the unknown state in order to reconstruct it further. \emph{Quantum State Tomography} (QST) is an experimental procedure where the ensemble of unknown, but identically prepared quantum states, is characterised by a sequence of measurements in different bases, enabling the reconstruction of its density matrix. Analogous to medical tomographic reconstructions, in QST, several measurements in the various bases are taken and combined to give a reconstruction of the complete (initial) quantum state.

Measurement on a quantum system generally gives a probabilistic result, and the measurement outcome only provides limited information about the state of the system, even when an ideal measurement device is used. QST consists of only finite measurements and the use of appropriate estimation algorithms. Hence, choosing optimal measurement sets, and designing efficient state reconstruction algorithms, are two critical issues in quantum state tomography.

Thus, we summarise the process of QST in Algorithm \ref{algo1}.   
\begin{algorithm}
    \caption{Quantum State Tomography}
    \label{algo1}
    \begin{algorithmic}[0]
        \State \textbf{Input:} Tolerance $\epsilon\ll 1$, create $n\in\mathbb{N}$ identical copies, $\left\{\ket{\phi_{i}}\right\}_{i=1}^{n}$, of the target state to be estimated $\ket{\psi}$ to form an ensemble
        \State \textbf{Initialize:} Parameter information $\mathbf{M}$, and density matrix $\rho$
        \While{$||\mathbf{M}-\rho||>\epsilon$}
            \For{each copy $\left\{\ket{\phi}_{i}\right\}_{i=1}^{n}$}
                \State Perform measurements on each state in the ensemble, forming a complete basis
                \State $\sum_{i=1}^{n}\ket{\phi_{i}}\bra{\phi_{i}}=I$
                \State Apply an estimation method, $\mathcal{E}$, to recover parameter information, $\mathbf{M}$, from the measurement results
                \State Calculate density matrix $\rho$
            \EndFor
        \EndWhile
        \State \textbf{Return:} Density matrix $\rho$
    \end{algorithmic}
\end{algorithm}

Once the density matrix is obtained for the reconstructed state, one needs to test it for physical viability, i.e. if the resulting density matrix is positive semidefinite (non-negative eigenvalues), and has unit trace. If these conditions are met, then such an estimation is correct.

QST is widely used in several application areas including Quantum Error Mitigation \cite{ref6, ref31}, State Estimation \cite{ref26, ref27}, applications to qubit measurement and experimental reconstruction \cite{ref28, ref29}, qudit system reconstructions \cite{ref33}, and Photonics \cite{ref30}.

\textit{Quantum Machine Learning} (QML) is a rapidly developing field that combines the disciplines of Classical Machine Learning (ML) with Quantum Computing (QC). Since the advent of QC as a research track, it became natural to coalesce the two fields to try and attain a computational advantage over the implementation of ML applications. These include the design of QML algorithms that have a speed-up over their classical counterparts, reduced computational complexity and resources over ML algorithms, and a reduction in the usage of physical hardware. There exists a plethora of ML applications to QST, and thus, the application of QML to QST was an organic next step. Therefore, this paper claims no novelty in the application of QML to QST, however, the novelty presented lies in the widespread coverage of the methods, the structured approaches, and the high fidelity of the results obtained.

In this paper, we discuss some of the classical approaches employed for reconstruction and thereafter build on these principles to apply quantum algorithms based on NISQ-era \cite{ref25} devices that can be employed to achieve efficient QST. 

This paper is segmented as follows: 

   In \hyperlink{sec2}{Sec. 2.}, we provide a comprehensive literature review of important works in the field, and describe, in sufficient detail, the methods employed, and the novelties of each research piece.

   In \hyperlink{sec3}{Sec. 3.}, we discuss the classical approaches to state tomography by giving detailed descriptions of the theoretical underpinnings of each method.

   In \hyperlink{sec4}{Sec. 4.}, we discuss the mathematical, computational, and circuit architecture of the QST methods used. 

   In \hyperlink{sec5}{Sec. 5.}, we discuss the results of each method employed.

   In \hyperlink{sec6}{Sec. 6.}, we provide a conclusion to this research by discussing all that was done, all that was achieved, and future avenues of research to explore.
\section{Literature Review} \hypertarget{sec2}{}
In \cite{ref1}, Torlai \emph{et al} proposed a new method using Neural Networks (NNs) that can be used to effectively reconstruct quantum states for generalised highly-entangled states using simplistic, but limited in number, experimental data. For a vast array of quantum devices, namely: Adiabatic quantum simulators, higher-dimensional ion traps, highly-entangled quantum circuits, and ultra-cold atoms, the authors have shown that their method can be applied, in addition to the construction of quantifiers that could be difficult to measure directly. Lastly, and most importantly, it was shown that the proposed method was robust against noise.

In \cite{ref2}, Schmale \emph{et al} use Convolutional Neural Networks (CNNs) to reconstruct quantum states. In particular, a novel approach to accurately estimate observables from tomographic measurement data using a variational manifold represented by a CNN is presented. It was demonstrated that their proposed method outperformed analogous classical methods, namely the Maximum Likelihood Estimate (MLE), by up to an order of magnitude, by achieving high classical fidelities, and by reducing the error of estimation. Perhaps most important was the finding that their method scaled polynomially as the size of the system grew, making the method more versatile and able to handle larger systems with more complex quantum states.

In \cite{ref3},  Quek \emph{et al} propose a new algorithm called \emph{Neural Adaptive Quantum Tomography} (NAQT) that is modular, fast, and highly flexible. Using NNs, the algorithm resiliently optimises measurements and reconstructs quantum states with high accuracy, while being agnostic to the number of qubits involved, and the type of measurements used. Further, it was demonstrated that the NAQT algorithm outperformed the Adaptive Bayesian Quantum Tomography (ABQT) and Maximum Likelihood Quantum Tomography (MLQT) methods in terms of speed and accuracy; however, NAQT achieves a comparable reconstruction accuracy when benchmarked against ABQT and MLQT. The modularity of the algorithm lies in its ability to be seamlessly retrained, in a reasonable time, in order to suit the task at hand.

In \cite{ref4}, Koutn\'{y} \emph{et al} apply Deep Learning (DL) to QST. Using identically-prepared copies of the system, NNs were used to prepare the density matrix of the system from measurements. The novelty introduced in this paper was the concept of the positivity of the density matrix baked into the NN architecture. It was demonstrated that this method achieved state-of-the-art results in terms of speed when compared to the MLE and Semidefinite Programming (SDP) approaches in reconstructing the density matrix, and hence, estimating the initial state. In particular, this novel approach was roughly four orders of magnitude faster than the MLE and three orders of magnitude faster than the SDP. Lastly, it was shown that the results naturally extend to Quantum Information tasks.

In \cite{ref6}, Hai and Ho introduce a new algorithm called \emph{Universal Compilation} (UC) that aims to maximise the efficiency of the QST task, with a focus on quantum sensing and metrology as application domains. Several novelties were introduced, namely: 

\begin{enumerate}
\item[2.1.] The introduction of a new cost function 
\begin{equation*}
C(\boldsymbol{\theta})=d(\Psi,\Phi(\boldsymbol{\theta})),
\end{equation*}
  
based upon the Fubini-Study metric (distance measurement between two quantum states) \cite{ref7}, for the purposes of this study, it is given by
\begin{equation*}
d(\Psi,\Phi)=\cos^{-1}\left(||\bra{\Psi}\ket{\Phi}||^{2}\right),
\end{equation*}
where $\ket{\Psi}$ is the target (unknown) state, and $\ket{\Phi(\boldsymbol{\theta})}$ is the parameterised variational (reconstructed) state with parameters $\boldsymbol{\theta}=\left(\theta_{1}, \theta_{2},\ldots,\theta_{n}\right)\in\left[0,1\right]$. The associated optimisation problem is stated as
\begin{equation*}
\boldsymbol{\theta}^{*}=\underset{\theta_{1},\theta_{2},\ldots,\theta_{n}}{\arg\min}\;C(\boldsymbol{\theta}),
\end{equation*}
using a gradient-based scheme, with updates of the form $\boldsymbol{\theta}\leftarrow\boldsymbol{\theta}-\eta\boldsymbol{\nabla}_{\boldsymbol{\theta}}C(\boldsymbol{\theta})$, for $0\leq\eta\leq 1$.  
\item[2.2.] The versatility of UC on qubit and qudit systems. 
\item[2.3.] The UC algorithm can improve the accuracy of the QST tasks in the presence of noise, and other sources of errors. 
\end{enumerate}

In \cite{ref10}, the MLE, least squares, generalised least squares, positive least squares, thresholded least squares, and projected least squares (PLS) methods have been analysed, compared, and the computational efficiencies of the methods have been presented. The measurement scenarios have been realised with respect to Pauli bases, random bases measurements, and the covariant measurement. In addition, a complete set of simulation results have been shared online via an interactive \texttt{Shiny} application.

In \cite{ref11}, a least-square inversion method, which reconstructs the density matrix from measurable time-dependent probability distributions of a physical system from oscillators (harmonic and anharmonic), has been proposed. The applicability of the method has been compared with other methods based on least-squares inversion.

The \emph{Quantum Process tomography} (QPT) task, which estimates unknown quantum transformations completely from measurement data, is a powerful, resource-intensive method employed in the field of quantum technology. In \cite{ref12}, the PLS method has been proposed and investigated for QPT. The PLS method computes the least-squares estimator of the Choi matrix \cite{ref32} of an unknown channel, and projects it onto the convex set of Choi matrices. The PLS method has been illustrated with numerical experiments involving channels on systems with up to a $7$-qubit system.

In \cite{ref13}, a new quantum algorithm, which determines the quality of a least-squares fit over an exponentially large dataset, used to solve a system of linear system equations efficiently, has been proposed. The proposed algorithm efficiently finds a concise function that approximates the data to be fitted and gives a bound for the approximation error. For pure quantum states, the algorithm performs an efficient parametric estimation of the quantum state. A use-case of this approach is the performance of a complete QST analysis.

In \cite{ref15}, a QML technique for QST on an unknown quantum state is introduced. Mechanically, the learning process of the technique involved maximising the fidelity between the output of a VQC and the state. The viability of the method has been demonstrated by performing numerical simulations for the tomography of the ground state of a one-dimensional quantum spin chain using a VQC simulator.

In addition to the methods described above, there have been several other ML approaches: Using conditional Generative Adversarial Networks (GANs) -- see \cite{ref8}, using attention mechanisms \cite{ref9}, and the ML literature contained there in the abovementioned research pieces. 
\section{Classical Methods for Quantum State Tomography}\hypertarget{sec3}{}
In this section, we present the theoretical foundations of the various classical techniques commonly used for quantum state tomography.
\subsection{Linear Inversion} 
The first method to appear in literature was \emph{linear inversion} \cite{ref22}, which is based on the inverse of Born's rule. The method entails equating experimentally normalised frequencies to the probabilities predicted from Quantum Mechanical calculations. The reconstructed state, however, might not match a physical state because of experimental noise. The predicted density matrix must be confined to the space of semidefinite, positive, and unit trace matrices in order to confirm its physicality. The statistical inference problem at hand can be described as being constrained parameter estimation.

Consider the $d$-dimensional Hilbert space, $\mathcal{H}^{d}$. Let $\Gamma_\mu$ represent elements of an orthonormal basis for Hermitian matrices in this space. 
Thus, 
\begin{equation}
\Tr\left(\Gamma_\mu \Gamma _\nu\right)=\delta_{\mu \nu},
\end{equation} 
where $\delta_{\mu\nu}$ is the Kronecker-delta function. The density matrix can then be written as
\begin{equation}
\rho = \sum_{\mu} S_\mu \Gamma_\mu,
\end{equation}
 where $S_\mu$ are also called the Stokes parameters for this representation. Thus, measurements described by Positive Operator-Valued Measure (POVM) set elements, {$O_\mu$}, yield results given by $ f_\mu = \Tr\left(O_\mu \rho\right)$.
These frequencies, in terms of the basis operators, are then given by
\begin{equation}
f_\mu=\sum_{\nu} S_\nu \Tr\left(O_\mu \Gamma_\nu\right). \label{3}
\end{equation}
Eq.\eqref{3} can we write vectorially as the linear equation
\begin{equation}
\mathbf{B}\mathbf{s}^{T}=\mathbf{f},  \label{4}  
\end{equation}
where $\mathbf{s}$ is the Stokes vector, $\mathbf{B}$ is the matrix calculated in terms of the Basis ${\Gamma_\nu}$ and the measurement operators ${O_\mu}$ as $ B_{\mu\nu}= \Tr\left(O_\mu \Gamma_\nu\right)$, and $\mathbf{f}$ is the frequency vector. Thus, this technique is termed the \emph{linear tomography reconstruction} or the \emph{linear inversion} method.

Solving the system of linear equations from Eq.\eqref{4}, however, is not guaranteed to yield a physical quantum state $\rho $ through the parameters $\mathbf{s}$. This is because real-world measurements have statistical noise, which leads to the solution to this equation defying the properties of unit trace and positive semi-definiteness.

The major drawback of this method is that an exponential number of measurements, with respect to the number of qubits in the system, is required to be performed. Similarly, the MLE methods suffer from this disadvantage; a discussion is left for the next section. 
\subsection{Maximum Likelihood Estimation}
This approach was developed by \v{R}eh\'{a}\v{c}ek \emph{et al} \cite{ref23} for discrete-variable systems. It has the following basic structure: Consider a system of $N$ identically prepared unknown quantum states. By using the results of the measurements made on this system, the identity of this unidentified quantum system is determined as defined by the density operator. The measurements produce multinomial statistics for the number of occurrences $n_j$ for each result $j$. The likelihood for this measurement data $\mathcal{D}$ is calculated.

Additionally, the Hermitian operator,
\begin{equation}
R=\sum_j \frac{f_j}{p_j}\Pi_j \label{5}
\end{equation}
was introduced as possessing the property of being positive semi-definite. This property encompasses the outcomes of the measurements and the corresponding measurement operators. In \eqref{5}, $f_j$ are the measured frequencies for outcome $j$, $\Pi_j$ are the corresponding measurement operators, and $p_j$ are the outcome expected from the quantum state $\rho$, i.e. $ p_j = \Tr\left(\rho\Pi_j\right)$.

Using $R$, one can find the reconstructed state, $\rho$, according to to the iterative equation 
\begin{equation}
\rho^{k+1} = \alpha R\rho^kR,
\end{equation}
where $\alpha$ is a normalisation constant that ensures that $\rho$ has unit trace. Note that $R$ here is not constant across all iterations, but itself depends on $\rho$.
The starting state $\rho^0$ for this procedure is taken to be the maximally-mixed state. i.e, $\rho^0 = I$.

The essence of classical MLE is the form of the loss function $ E(\rho) $, given by
\begin{equation}
E(\rho)=\sum_j f_j \log\left[\Tr\left(\rho\Pi_j\right)\right]. \label{7}
\end{equation}
Minimisation of Eq.\eqref{7}, by the standard gradient descent algorithm, for a particular parameterisation of $\rho$ can be shown to be equivalent to the iterative approach that is mentioned in \cite{ref21}.

To ensure positivity and unit trace the parameterisation employed is of the following form.
\begin{equation}
\rho=\frac{\mathcal{T}\mathcal{T}^{\dagger}}{\Tr\left(\mathcal{T}\mathcal{T}^{\dagger}\right)},\label{8}
\end{equation}
where $\mathcal{T}$ is taken to be of upper triangular form with complex entries.
\subsection{Least Squares Estimation} \hypertarget{sec3.3}{}
In this section, the details of the Least Square Estimation (LSE) for the QST are presented. The idea behind Least Square Estimation traces its roots back to Gauss in 1809 when studying the motions of celestial bodies in conic sections around the sun. Today it is a standard tool in Statistics and supervised ML for a profusion of applications, and very well known. Below, we outline the mechanics of the method in brief \cite{ref17, ref18}.

LSE finds the best linear fit for which the summation of the squares will be minimum. To minimise the error value, we take the first derivative equation.

Suppose that you have $n$ samples of data $\left(X(1),X(2),\ldots,X(n)\right)$, and the corresponding output labels for this data $\left(Y_i(1), Y_i(2),\ldots,Y_i(n)\right)$.
\begin{equation}
\mathbf{X} = \left[X_1, X_2,\ldots,X_{nq}\right]^T,
\end{equation}
where $\mathbf{X}$ is the vector representation of the features. We assume that output is linearly related to the input, and for the $i^{\text{th}}$ data point, the equation is
\begin{equation}
Y_i(k) = Q_1 X_1(k) + Q_2 X_2(k)+\ldots+Q_{nq} X_{nq}(k)+E(k), \label{10}
\end{equation}
where $E(k)$ denotes the error value. This is generalised in, matrix form, to
\begin{equation}
\mathbf{Y}(k) = \mathbf{X}^T(k)\mathbf{Q}(k) + \mathbf{E}(k), 
\end{equation}
where
\begin{equation*}
\mathbf{X} =
\begin{bmatrix}
X_1(1) &X_{2}(1) & \dots & X_{nq}(1) \\
X_{1}(2) &X_{2}(2) &\ldots &X_{nq}(2) \\
\vdots &\vdots &\ddots & \vdots \\
X_1(n) &X_{2}(n) & \dots & X_{nq}(n)
\end{bmatrix}.
\end{equation*}
The goal is to estimate the values of $\mathbf{Q}$ by minimising the sum of the squares of the error values. The associated cost function, known as the \emph{Mean Squared Error} (MSE) cost, denoted $J(\mathbf{Q})$, is given by
\begin{equation}
\begin{split}
J(\mathbf{Q})=&\; \frac{1}{2} \mathop{\sum} \limits_{k=1 }^n {E^2 (k)} = \mathbf{E}^T\mathbf{E}\\
=&\; \frac{1}{2} \left(\mathbf{Y}-\mathbf{X}\mathbf{Q}\right)^T\left(\mathbf{Y}-\mathbf{X} \mathbf{Q}\right)\\
=&\; \frac{1}{2} \left(\mathbf{Y}^T-\mathbf{X}^T \mathbf{Q}^T\right)\left(\mathbf{Y}-\mathbf{X}\mathbf{Q}\right). \label{12}
\end{split}
\end{equation}
Taking the derivative of Eq.\eqref{12}, we obtain
\begin{equation}
\frac{\delta J(\mathbf{Q})}{\delta \mathbf{Q}}=-\mathbf{Y}^T\mathbf{X}+\mathbf{Q}^T\mathbf{X}^T\mathbf{X}. \label{13}
\end{equation}
Setting Eq.\eqref{13} to zero, and solving for $\mathbf{Q}$, we obtain
\begin{equation}
\mathbf{Q} = \left(\mathbf{X}^T \mathbf{X}\right)^{-1}\mathbf{X}^T \mathbf{Y}.
\end{equation}
\subsection{ML and Covariance Matrix Estimation}
We provide a brief review of the developments in estimating large covariance matrices and precision matrices. In order to estimate these large covariance matrices, the assumption for the target matrix of interest is sparsity which is most commonly the case for precision matrices \cite{ref19}. There are two general approaches, 
\begin{enumerate}
\item[3.4.1.] \textbf{Rank-based Approach:} This technique is applicable when the processing of data exhibits a heavily-tailed, or non-Gaussian distribution.
\item[3.4.2.] \textbf{Factor-model-based Approach:} To deal with conditional sparsity, the remaining elements of the output variables become sparse. For this, a factor model is required. 
\end{enumerate}

Covariance estimation uses factor models by utilising the principal components method. The most common approach to the estimation of sparse precision matrices is the MLE. Let $Y_{1},Y_{2},\ldots,{Y_T}\overset{\text{i.i.d.}}{\sim}\mathscr{D}$ be random variables that are distributed independently and identically, then the negative Gaussian log-likelihood is defined as 
\begin{equation}
G(\Theta) = \Tr(S\Theta) -\log|\Theta|,
\end{equation}
where, for independently weak data and non-Gaussian data, $G(\Theta)$ behaves as a quasi-negative log-likelihood. We consider the penalised likelihood approach
\begin{equation}
\hat{\Theta} = \mathop{\arg \min}\limits_{\Theta = (\theta_{ij})_{pxp}} {\Tr\left(S\Theta\right) - \log|\Theta|} + \mathop{\sum} \limits_{i \neq j } P_{w_{T}}\left(|\theta_{ij}|\right),
\end{equation}
where $P_{w_{T}}$ is the penalty function that encourages the sparsity of $\hat{\Theta}$. The penalties that are commonly used are convex penalties, i.e., L1-penalty (lasso). However, in contrast, concave penalties tend to perform better. As the value of the parameter increases, it minimises the shrinkage bias of folded concave penalties.


Thereafter, we look to solve the adjacent problem of estimating the covariance matrix \(\boldsymbol{\Sigma}\) under the elliptical model, which poses a significant challenge, especially when sparse estimations are required. Adopting a similar approach as the previous section, we impose a sparsity assumption on \(\boldsymbol{\Sigma}\). To address this issue, the EC2 (\textbf{E}stimation of \textbf{C}ovariance with \textbf{E}igenvalue \textbf{C}onstraints) estimator is used, which is a regularised rank-based estimation method. This is regarded as a natural extension of the generalised thresholding operator.

An exceptional feature of the EC2 estimator is its ability to guarantee the positive definiteness of the estimated covariance matrix. This is achieved by explicitly constraining the smallest eigenvalue of the estimated covariance matrix, a feature not present in many existing methods.

The EC2 estimator is defined for the sparse covariance matrix as
\begin{equation}
\widehat{\mathbf{R}}^{\mathrm{EC} 2}:=\underset{\operatorname{diag}(\mathbf{R})=1}{\operatorname{\arg\min}} \frac{1}{2}\|\widehat{\mathbf{R}}-\mathbf{R}\|_{\mathrm{F}}^{2}+\lambda\|\mathbf{R}\|_{1, \text { off }},
\end{equation}
such that $\tau \leq \Lambda_{\min }(\mathbf{R})$. In this equation, \(\widehat{\mathbf{R}}^{\mathrm{EC} 2}\) represents the sparse estimator for \(\mathbf{R}\), derived by simultaneously imposing a sparse estimation and a positive-definiteness constraint. The regularisation parameter, \(\lambda>0\), together with the desired minimum eigenvalue lower bound, \(\tau>0\), are predetermined. The constraint \(\operatorname{diag}(\mathbf{R})=1\) ensures the resulting \(\widehat{\mathbf{R}}^{\mathrm{EC} 2}\) is a correlation matrix. To finalise the process, the covariance matrix estimator, \(\widehat{\boldsymbol{\Sigma}}\), is obtained by converting \(\widehat{\mathbf{R}}^{\mathrm{EC} 2}\).

The EC2 estimator's asymptotic properties are crucial in understanding its behaviour. We consider two classes:
\begin{enumerate}
\item The class of sparse correlation matrices:

\begin{multline}
\mathcal{M}(q, M_{p}, \delta) := \biggl\{
\mathbf{R} : \max_{1 \leq j \leq p} \sum_{k \neq j} |\mathbf{R}_{jk}|^{q} \leq M_{p} \\
\text{and } \mathbf{R}_{jj} = 1 \text{ for all } j, \Lambda_{\min}(\mathbf{R}) \geq \delta
\biggr\}.
\end{multline}
\item The class of covariance matrices:
\begin{equation}
\begin{aligned}
\mathcal{U}(\kappa, q, M_{p}, \delta)
& := \left\{
\boldsymbol{\Sigma} : \max_{j} \boldsymbol{\Sigma}_{jj} \leq \kappa
\right. \\
& \text{ and } \mathbf{D}^{-1} \boldsymbol{\Sigma} \mathbf{D}^{-1} \in \mathcal{M}(q, M_{p}, \delta)
\left. \vphantom{\max_{j}} \right\}.
\end{aligned}
\end{equation}

\end{enumerate}

Given an elliptical distribution assumption, the EC2 estimator, \(\widehat{\Sigma}\), is bound by
\begin{equation}
\sup _{\boldsymbol{\Sigma} \in \mathcal{U}\left(\kappa, q, M_{p}, \delta_{\min }\right)} \mathbb{E}\left\|\widehat{\Sigma}^{\mathrm{EC} 2}-\boldsymbol{\Sigma}\right\|_{2} \leq c_{1} \cdot M_{p}\left(\frac{\log p}{T}\right)^{\frac{1-q}{2}},
\end{equation}
where \(T\) represents the sample size and \(c_{1}\) is a constant. Notably, the EC2 estimator attains the minimax lower bound over the class \(\mathcal{U}\left(\kappa, q, M_{d}, \delta_{\min }\right)\) in the Gaussian model scenario, rendering it asymptotically rate optimal.

We can use the resulting covariance matrix as our density matrix and perform Principal Components Analysis (PCA) to reduce the dimensions needed to describe our original quantum state \cite{ref43}.
\subsection{Bayesian Methods}
\emph{Bayesian Inference} is a parameter estimation method that guarantees a positive semidefinite density matrix with unit trace and is robust against noise \cite{ref16}.

Bayesian inference hinges on the Bayes theorem from probability theory, which asserts that the probability of estimating a parameter $\theta$ given some data, $\mathcal{D}$, can be calculated by
\begin{equation}
\label{eq:bayes}
\mathbb{P}\left(\theta | \mathcal{D} \right)= \frac{\mathbb{P}\left( \mathcal{D} | \theta \right) \mathbb{P}\left(\theta\right)}{\mathbb{P}\left(\mathcal{D}\right)},
\end{equation}
where $\mathbb{P}\left(\theta | \mathcal{D} \right)$ is called the \emph{posterior probability}, $\mathbb{P}\left( \mathcal{D} | \theta \right)$ is the ``likelihood of obtaining said data, $\mathcal{D}$'', $\mathbb{P}\left(\theta\right)$ is the prior probability, and $\mathbb{P}\left(\mathcal{D}\right)$ is the \emph{evidence} \cite{ref17}.

Given a complete set of observables $\{E_j\}_{j=1}^M$, where $M$ is restricted by $M\geq d^2-1$. For $N$ identically prepared states, the probability distribution of getting a particular number of specific outcomes given a density matrix $\rho$ (i.e. the likelihood) is given by:
\begin{equation}
\label{eq:likelihood}
\begin{split}
\mathbb{P}\left(\textbf{N}=D | \rho\right)&=\mathbb{P}\left(N_1=n_1,\ldots,N_M=n_M |\rho\right)\\
&=\frac{N!}{\prod_{i=1}^M n_i!}\prod_{j=1}^Mp\left(j|\rho\right)^{n_j},
\end{split}
\end{equation}
where $D=(n_1,n_2,\ldots,n_M)$ is known as the data vector which contains the results of our measurement experiments, and $N_j$ is the number of times we get outcome $j$ in our $N$ trials. From Born's rule,
\begin{equation}
p\left(j | \rho\right)=\Tr\left(\rho E_j\right).
\end{equation}
In order to estimate a density matrix $\rho$, we first need to parameterise it. One way of doing this is by writing it in terms of the Cholesky decomposition in Eq.\eqref{8}. Additionally, we note the form of $\mathcal{T}(t)$
\begin{equation*} 
\mathcal{T}(t)=
\begin{bmatrix}
t_1 & t_{d+1}+it_{d+2} &\dots & t_{d^2-1}+it_{d^2} \\
0 & t_2 & \dots & \vdots\\
0 & 0 & \ddots & \vdots \\
\vdots & \ddots & \ddots & t_{3d-3}+it_{3d-2} \\
0 & \dots &\dots & t_d 
\end{bmatrix}, 
\end{equation*}
where $d$ is the dimension of the density matrix $\rho$ which we want to estimate. In order to ensure that the log-likelihood remains concave, the following restrictions are put on the parameters $t_i$ \cite{ref18}:
\begin{enumerate}
\item \textbf{Unit length:} $||t_i||_2^2=1$.
\item \textbf{Strict positivity:} $t_i>0$.
\end{enumerate}
Further, changing the parameter space from $t_i$ to $\theta_i$ is desirable to enforce the above restrictions. This is achieved through the transformation
\begin{equation}
t_i=\cos{\theta_{i}}\prod_{j=i}^{d^2}\sin{\theta_j}.
\end{equation}
Finally, we can estimate our parameters in Eq.\eqref{eq:bayes} 
\begin{equation}
\label{eq:analytic}
\pi\left(\theta | \mathcal{D}\right) \propto \frac{N!}{\prod_{i=1}^M n_i!}\prod_{j=1}^M \Tr\left[\rho(\theta)E_j\right]^{n_j} \pi(\theta).
\end{equation}
A careful choice of the prior probability, $\pi\left(\theta\right)$, must be made. As one may notice, Eq.\eqref{eq:analytic} is still not tractable analytically, so numerical techniques like the MCMC Metropolis-Hastings algorithm \cite{ref16} can be used for this purpose.

\section{Quantum Machine Learning Methods for Quantum State Tomography}\hypertarget{sec4}{}
In this section, we present the theory of the various QML methods used in this study. 

\subsection{Variational Quantum Circuit Algorithm}

\emph{Variational Quantum Circuits} (VQCs) are widely used in QML tasks for solving problems via parameter updates which are consistent with minimising the loss function \cite{ref21, ref39}. The methodology of this algorithm can be summarised as follows: 
\begin{enumerate}
\item[4.1.1.] \textbf{State Preparation:} Encode the classical data into quantum states. 
\item[4.1.2.] \textbf{Parameter Adjustment:} Finding the optimal values of your parameters such that the loss function is a minimum.
\item[4.1.3.] \textbf{Measurement:} Measuring on a classical computer. 
\end{enumerate}
As this circuit's parameters increase, the number of qubits also increases polynomially. The circuit approaches highly entangled states with these polynomial parameters. 
In this circuit, half of the qubits are used to indicate the mixed target state, which is obtained by QST and denoted as a $\widehat{\rho }$, and the other half of the qubits are used as auxiliary qubits. The output of the circuit is denoted as $\mathbf{|\Psi \rangle}$ as expressed in the Eq.\eqref{eq:vqc1}
\begin{equation}
|\Psi \rangle = V(\boldsymbol{\theta}) |0 \rangle^{\otimes 2n}.
\label{eq:vqc1}
\end{equation}
Tomography of the first $n$-qubits, which is used for the mixed target state, is obtained by Eq.\eqref{eq:vqc2}
\begin{equation}
{\widehat{\rho}}^T = \Tr\left(|\Psi \rangle \langle \Psi |\right). 
\label{eq:vqc2}
\end{equation}
The fidelity and loss function are calculated using Eqs.\eqref{eq:vqc3} and \eqref{eq:vqc4}, respectively. The purpose is to maximise the value of fidelity while minimising loss value.
\begin{align}
A(\boldsymbol{\theta})=&\;\Tr\left({\widehat{\rho}} {\widehat{\rho}}^T\right)=\left(\langle \Psi|{\widehat{\rho}} {\otimes}I|\Psi \rangle\right), \label{eq:vqc3} \\
a(\boldsymbol{\theta})=&\;1-\sqrt{(A(\boldsymbol{\theta})}.\label{eq:vqc4}
\end{align}

\begin{figure}[h!]
\centering
\includegraphics[width=0.75\linewidth]{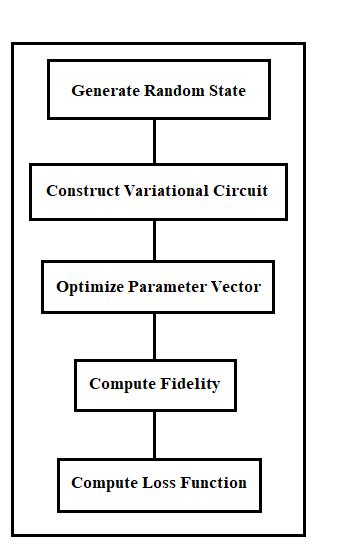}
\vspace{-0.5cm}
\caption{Flowchart of the Variational Quantum Circuit for QST Algorithm.}
\label{fig:vqc1}
\end{figure}

\begin{figure*}[t]
\centering
\includegraphics[width=1\linewidth]{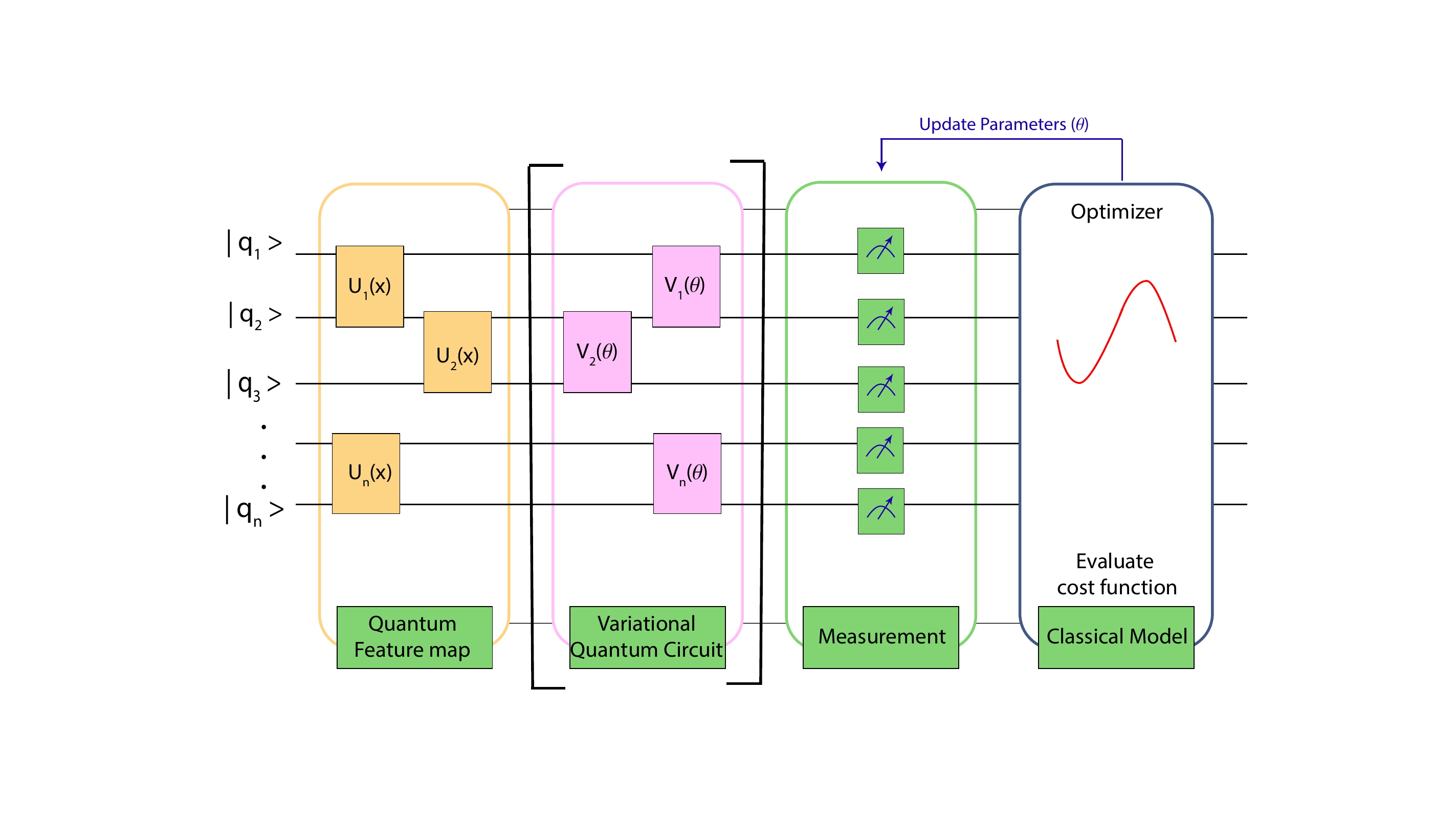}
\vspace{-2cm}
\caption{Variational Quantum Circuit diagram. The VQC initiates with $n$ input qubits that undergo a feature map before being processed through the circuit itself. After this transformation, all qubits are measured, and their outcomes are subsequently assessed by a classical model to evaluate the cost function.}
\label{fig:vqc2}
\end{figure*}

A flowchart of the VQC for implementation in the QST algorithm has been illustrated in Fig~\ref{fig:vqc1}. This algorithm uses a classical quantum framework, as shown in Fig~\ref{fig:vqc2}.

In the first step of the VQC for the QST algorithm, the \texttt{GenerateRandomPsi} function is called to initialise the target state. This function gets the number of qubits, determined as $5$ in this implementation, and returns the \texttt{statevector} object using the \texttt{RandomStatevector} library. Once the $\boldsymbol{\theta}$-parameter vector is initialised, the algorithm calls the \texttt{InitializeTheta} function, which gets the \texttt{CircDepth} and \texttt{NumQbits} as parameters and returns $\boldsymbol{\theta}$ values. 

\begin{figure*}[h!]
\centering
\includegraphics[width=\linewidth]{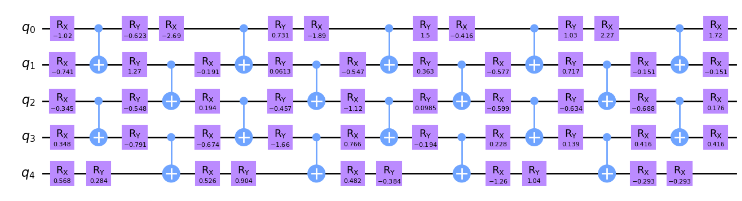}
\vspace{-0.5cm}
\caption{Variational Quantum Circuit used for QST. The depicted quantum circuit is constructed using the \texttt{Construct VariationalCirc} algorithm, employing rotational \texttt{Rx}, \texttt{Ry}, and \texttt{CNOT} gates to represent generic quantum states. Each layer consists of \texttt{CNOT} gates, defining the circuit depth $s$, which is set to 10 layers for this representation.}
\label{fig:vqc3}
\end{figure*}

In the second step of the algorithm, \texttt{Construct VariationalCirc} is called to generate a parameterised VQC as seen in Fig~\ref{fig:vqc3}. To represent generic quantum states, we used rotational \texttt{Rx}, \texttt{Ry}, and \texttt{CNOT} gates. Each layer of this circuit with \texttt{CNOT} gates is determined as a depth $s$, and it is set to $10$ in this step.

In the third step, the objective is to minimise the discrepancy between the predicted and target states, which is expressed by the loss function. Notably, this loss function is equivalent to the LSE from \hyperlink{sec3.3}{Sec. 3.3.} 
To achieve this, the algorithm invokes the \texttt{OptimizeThetaScp} function, responsible for ascertaining the optimally parameterised vector denoted as $\mathbf{\theta}$.

To effectuate this optimisation objective, the algorithm employs the \texttt{ComputeLossGradient} function, which calculates the gradient of the loss function \cite{ref39}. This gradient facilitates the iterative refinement of the parameter vector $\boldsymbol{\theta}$ in accordance with Eq. \eqref{eq:vqc5}:
\begin{equation}
\frac{\delta a(\boldsymbol{\theta})}{\delta \boldsymbol{\theta} _i}=\frac{\delta a(\boldsymbol{\theta})}{\delta A(\boldsymbol{\theta})} \frac{\delta A(\boldsymbol{\theta})}{\delta \boldsymbol{\theta}}. 
\label{eq:vqc5} 
\end{equation} 

In the fourth step, the fidelity between the predicted and target quantum states is calculated using the \texttt{ComputeFidelity} function from the \texttt{StateFidelity} library.

In the final step, the loss is calculated using the fidelity values obtained from the preceding step.
\subsection{Quantum Principal Components Analysis Algorithm}
In this section, we consider using Quantum Principal Component Analysis (qPCA) in lieu of classical PCA to achieve computational speed-ups and efficient density matrix estimation. 

The qPCA algorithm proposed by Lloyd \textit{et al} \cite{ref36} states that $n$ copies of a quantum system with a density matrix $\rho$ in a $d$-dimensional Hilbert space can be employed to implement the unitary operator $U = e^{-\imath \rho t}$. This is based on the idea that the density matrix, $\rho$, can function analogous to a Hamiltonian operator, generating transformations on other states. The idea that multiple copies of $\rho$ are used to construct a unitary transformation applied to $\rho$ demonstrates that a quantum state can play a dynamic role in its own analysis. Instead of the state being a passive entity upon which measurements are made, it actively transforms other states.

The following relation serves as the foundation for the mathematical framework of qPCA, where the unitary $U = e^{-\imath \rho t}$ is applied to any density matrix $\sigma$ up to the $n^{\text{th}}$ order in $t$ 
 \begin{equation}
 \operatorname{Tr}_{P} e^{-\imath S \Delta t} \rho \otimes \sigma e^{\imath S \Delta t} = \sigma - \imath \Delta t[\rho, \sigma] + \mathcal{O}(\Delta t^{2}),
 \label{eq:qpca1}
\end{equation}
  where $\operatorname{Tr}_{P}$ is the partial trace over the first variable, $S$ is the swap operator, and $\rho$ and $\sigma$ are density matrices. Repeated application of the above equation, with $n$ copies of $\rho$, facilitates the construction of $e^{-\imath \rho n \Delta t} \sigma e^{\imath \rho n \Delta t}$. Simulating $e^{-\imath \rho t}$ to an accuracy of $\varepsilon$ demands $n = \mathcal{O}(t^2 \varepsilon^{-1}||\rho - \sigma||_{\infty}^2) \leq \mathcal{O}(t^2 \varepsilon^{-1})$ steps, where $t = n \Delta t$ and $||\ldots ||_{\infty}$ is the sup norm.

 Having $ n $ copies of $ \rho $ allows applying the unitary transformation $ e^{-\imath \rho t} $ to perform the quantum phase algorithm. This algorithm can transform any initial state $ |\psi\rangle|0\rangle $ to
\begin{equation}
\sum_{i} \psi_{i}|\chi_{i}\rangle|\tilde{r}_{i}\rangle,
 \label{eq:qpca2}
\end{equation}
where $|\chi_{i}\rangle$ are the eigenvectors of $\rho$ and $\tilde{r}_{i}$ are estimates of the corresponding eigenvalues, and $\psi_i = \braket{\chi_i}{\psi}$. 
Given $ t = n \Delta t $, consider the unitary can be expressed as:
\begin{equation}
\sum_{n}|n \Delta t\rangle\langle n \Delta t| \otimes \prod_{j=1}^{n} e^{-\imath S_{j} \Delta t}. 
\label{33}
\end{equation}
When Eq.\eqref{33} is applied to the state as expressed in
\begin{equation}
|n \Delta t\rangle\langle n \Delta t| \otimes \sigma \otimes \rho \otimes \ldots \otimes \rho,
 \label{eq:qpca3}
\end{equation}
with $ \sigma = |\chi\rangle\langle\chi| $ and $ S_{j} $ being the swap operator for the $ j^{\text{th}}$ copy of $ \rho $, taking the partial trace over the copies of $ \rho $ provides the desired transformation
\begin{equation}
|t\rangle|\chi\rangle \mapsto |t\rangle e^{-\imath \rho t}|\chi\rangle. 
\end{equation}
 
 Integrating this into the quantum phase algorithm and using enhanced phase estimation techniques allows for the precise extraction of eigenvectors and eigenvalues of $ \rho $ in time $ t = \mathcal{O}(\varepsilon^{-1}) $, demanding $ n = \mathcal{O}(1/\varepsilon^{3}) $ copies of $ \rho $.
 
Applying the quantum phase algorithm with $ \rho $ as the initial state yields
\begin{equation}
\sum_{i} r_{i} |\chi_{i}\rangle\langle\chi_{i}| \otimes |\tilde{r}_{i}\rangle\langle\tilde{r}_{i}|.
\end{equation}
Sampling from this state unveils information about the eigenvectors and eigenvalues of $ \rho $.

\begin{algorithm}[t]
\caption{qPCA for QST}
\label{algo:qpca_algorithm}

\textbf{Input:} Multiple copies of a quantum state $\rho$, desired accuracy $\epsilon$.\\
\textbf{Output:} Eigenvectors and eigenvalues of $\rho$ in quantum form. \\
\textbf{Step 1: Initialise the parameters} 

\textbf{Step 2: Density Matrix Exponentiation} \\
Use $n$ copies of $\rho$ to apply unitary transformations of the form $e^{-\imath \rho t}$.

\textbf{Step 3: SWAP Operations} \\
Use repeated infinitesimal swap operations on $\rho \otimes \sigma$ to construct the unitary operator $e^{-\imath \rho t}$.

\textbf{Step 4: Quantum Phase Algorithm} \\
Use the ability to apply $e^{-\imath \rho t}$.\\
Perform the quantum phase algorithm to take any initial state $|\psi\rangle|0\rangle$ to a superposition of the form $\sum_{i} \psi_{i}|\chi_{i}\rangle|\tilde{r}_{i}\rangle$.

\textbf{Step 5: Conditional Operation} \\
Implement the conditional operation by replacing the SWAP operator with a conditional SWAP.

\textbf{Step 6: Eigenvector and Eigenvalue Decomposition} \\
Use the quantum phase algorithm with $\rho$ as the initial state to obtain the state $\sum_{i} r_{i}|\chi_{i}\rangle\langle\chi_{i}|\otimes| \tilde{r}_{i}\rangle\langle\tilde{r}_{i}|$.

\textbf{Step 7: Sampling} \\
Sample from this state to reveal features of the eigenvectors and eigenvalues of $\rho$.

\textbf{Step 8: Output} \\
Return the eigenvectors and eigenvalues in quantum form.
\end{algorithm}
 
qPCA proves invaluable when $\rho$ has a small rank, $ R $, or can be approximated with rank $R$. In such scenarios; only the largest $R$ eigenvalues will be non-zero in the eigenvector/eigenvalue decomposition. If $m\times n$ copies of $\rho$ are used, $m$ copies of this decomposition are obtained. The eigenvalues can then be analysed by quantum measurements on their corresponding eigenvectors for a chosen Hermitian operator $M$, provided $M$ is sparse or can be efficiently simulated. This allows the examination of the eigenvalues and eigenvectors of an unknown $\rho$ in time $\mathcal{O}(\log d)$.

Several improved qPCA algorithms include those that are based on quantum singular-value thresholding, that reduce the number of samples of measurement required \cite{ref41, ref42}, and others that reduce the number of ancillary qubits required for Quantum Phase Estimation (QPE) \cite{ref38}. Additionally, the algorithm by Lloyd \textit{et al} can be applied to classical data by representing the classical data as quantum states and constructing a covariance matrix, which is used as the density matrix \cite{ref35}. However, classical algorithms that are analogous to the qPCA algorithm by Lloyd \textit{et al}, using the $\ell^2$-norm sampling assumptions, suggest that the speed-ups from qPCA on classical data arise from its dependence on strong input assumptions, meaning qPCA is less useful on classical data when compared to analogous classical algorithms \cite{ref37}. Thus, the qPCA algorithm by Lloyd \textit{et al} might be more applicable for quantum data.

We then test how the density matrix $\rho_\text{reconstructed}$, which is reconstructed from the eigenvectors resulting from QPE, differs based on $t$. To do so, we introduce a modified algorithm based on (\ref{algo:qpca_algorithm}), wherein the density matrix of $\rho_\text{original}$ is known beforehand, and the Pauli strings of the state are provided. In this case, all that is required is the application of the unitary $U = e^{-\imath \rho t}$ in QPE. Here, we use QPE using the Quantum Fourier Transform \cite{ref40} with $2$ auxiliary qubits for eigendecomposition. Then, the resulting eigenvectors are used to construct $\rho_\text{reconstructed}$. However, because the dimensionality of $\rho_\text{reconstructed}$ is reduced when compared to that of $\rho_\text{original}$, we need to embed both states into a larger Hilbert space such that they have the same dimension. This can be done by taking the tensor product of the smaller state with an identity matrix of the appropriate size according to the following algorithm:
\begin{enumerate}
\item Given $\rho$ has dimension $d_1 \times d_1$ and $\sigma$ has dimension $d_2 \times d_2$, determine the larger dimension, $d = \max\left(d_1, d_2\right)$.
\item If $d_1 < d$, then take the tensor product of $\rho$ with the identity matrix of size $(d/d_1) \times (d/d_1)$. This would look like $\rho' = \rho \otimes I_{d/d_1}$.
\item If $d_2 < d$, then take the tensor product of $\sigma$ with the identity matrix of size $(d/d_2) \times (d/d_2)$. This would look like $\sigma' = \sigma \otimes I_{d/d_2}$.
\end{enumerate}
Subsequently, the fidelity, $F$, can be computed, where $F$ is calculated by the Uhlmann fidelity formula \cite{ref100}
\begin{equation}
F(\rho, \sigma) = \pars{\Tr \sqrt{\sqrt{\rho} \sigma \sqrt{\rho}}}^2.
\end{equation}

\subsection{Bayesian Approach}
In Bayesian Inference, probabilities are taken as a prior and post-prior distribution. The goal is to define and analyse these distributions. The unknown parameter $\theta$ is a prior distribution $\pi(\theta)$.

We propose the method for Bayesian QST below as explained in Algorithm \ref{algo:qst_bayesian}. The computational steps required are:
\begin{itemize}
\item Perform measurements on an unknown state $\rho$. This amounts to a total of $n$ individual outcomes.
\item Compute the LSE $\rho_{LS}$. If there are incomplete measurements, then $\rho_{LS}$ lives in subspace determined by only observed directions which can be represented by $\rho_{LS} = \bigl[P_{M}(\rho)\bigl]_{LS}$.
\item Parameterise the $N \times N$ density matrix by $N$ and non-negative real numbers $y_{N}\in\mathbb{R}\backslash\left(-\infty,0\right]$, and a column vectors $z_{N}$ of length $N$. The density matrix for the parameter set $\theta = \bigl[y_{1}, y_{2},\ldots,y_{N}; z_{1}, z_{2},\ldots,z_{N}\bigl]$ is written as,
\begin{equation}
\rho(\theta) = \mathop{\sum}\limits_{i = 1}\limits^{N} \left(\frac{y_{N}}{\sum\limits_ly_l}\right)\frac{z_{N}z_{N}^{\dagger}}{|z_{N}|^{2}}.
\end{equation}
\item Take the prior distribution for $\theta$ as
\begin{equation}
\pi_{0}{(\theta)} \propto \mathop{\Pi}\limits_{i = 1}\limits^{N} y_{N}^{a-1}e^{-y_{N}}e^{-\frac{1}{2} z_{N}^{\dagger}z_{N}},
\end{equation}
where $\alpha$ is a non-negative vector with scalar coefficients. $\pi_{0}(\theta)$ are random Gamma-distribution variables, and $z_{N}\sim\mathcal{N}(0,1)$ are standard normal Gaussian distributions.
\item The Likelihood function is given by
 \begin{equation}
\mathbb{P}(\mathcal{D}|\theta) \propto \exp\left[-\frac{N}{2} ||P_{M}(\rho(\theta)) - \rho_{LS}||_{F}^{2}\right],
\end{equation}
where $||P_{M}(\rho(\theta)) - \rho_{LS}||_{F}^{2}$ represents the Frobenius norm for the difference between the two matrices, $P_{M}(\cdot)$ is the projection operator, and $\rho_{LS}$ is the density matrix for estimating the least squares.
\item Make samples of invariant distribution $\pi(\theta) \propto \mathbb{P}(\mathcal{D}|\theta)\pi_{0}(\theta)$ as
\begin{equation}
\pi(\theta | \mathcal{D}) = \frac{1}{A}\times\mathbb{P}(\mathcal{D}|\theta)\pi_{0}(\theta), \label{post}
\end{equation}
where $A$ is the normalisation constant and is defined as $\int dx\;\pi(\theta) = 1$. Here $\mathbb{P}(\mathcal{D}|\theta)$ is the post-prior (posterior) probability with distribution of $\theta$, $\pi_{0}$ is the prior-distribution and $\mathbb{P}(\mathcal{D}|\theta)$ is the likelihood function.
\end{itemize}
Through these samples, we can estimate any function of $\rho$ using Eq.\eqref{post}.

\begin{figure*}[t]
\centering
\includegraphics[width=1\linewidth]{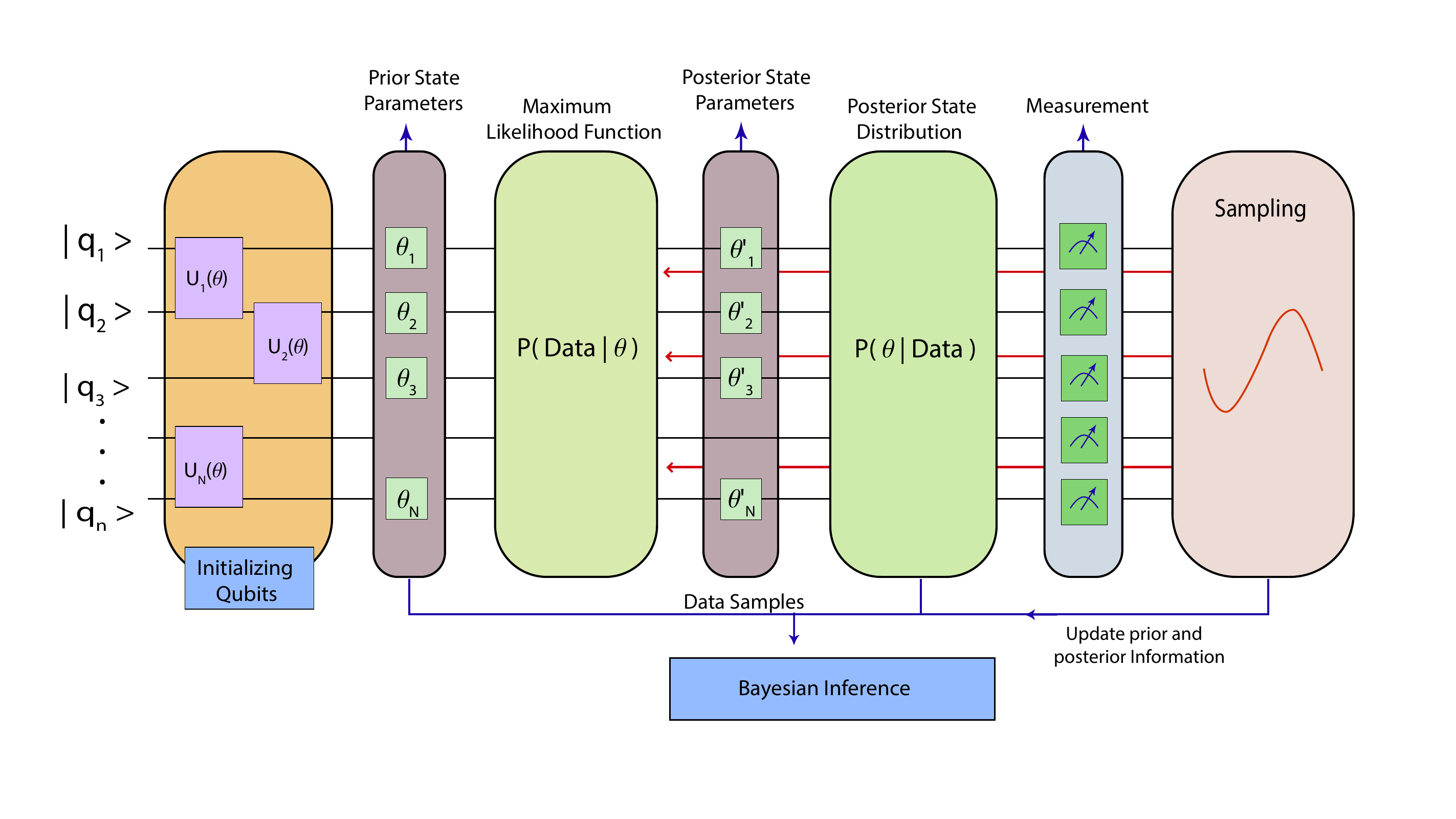}
\vspace{-1.5cm}
\caption{A schematic diagram of Bayesian Inference for Quantum State Tomography. $q$ defines the $Quantum Registers$ for $n$-qubits. While $\theta_1...\theta_n$ are prior parameters. $P(\mathcal{D}|\theta)$ represents a Maximum Likelihood function. $\theta_1'$ are posterior parameters used for Posterior State probability distribution, and then after measurements, the sampling procedure is performed.}
\label{fig:bqst}
\end{figure*}

\begin{algorithm}[t]
\caption{Bayesian Inference for QST}
\label{algo:qst_bayesian}
\textbf{Input:} Number of qubits, $N_{\text{qubits}}$, circuit depth, $D_{\text{circ}}$, prior and posterior parameters $\left(\boldsymbol{\theta},\boldsymbol{\theta}'\right)$ of the same length and size. \\
\textbf{Step 1: Initialise Parameters} \\
Define the density matrix of length dimensions $N$ for the parameterised quantum circuit as $\rho(\boldsymbol{\theta})$. \\
\textbf{Step 2: Prior Parameters} \\
Initialise the parameter vector $\boldsymbol{\theta}$ for generating samples of the prior distribution $\mathbb{P}(\mathcal{D})$.\\
\textbf{Step 3: Estimating Prior Sample Data} \\
Implement the log-likelihood Function $\mathbb{P}(\mathcal{D}|\boldsymbol{\theta})$ to estimate the sample data between the predicted and the target state based on given prior parameters. \\
\textbf{Step 4: Posterior Parameters} \\
Optimise the parameter vector $\boldsymbol{\theta}'$. \\
Perform a set of operations on the predicted and the target state.\\
\textbf{Step 5: Posterior Distribution} \\
Make samples of invariant distribution of vector $\boldsymbol{\theta}'$. \\ Compute the posterior probability distribution $\pi_{0}(\boldsymbol{\theta}')$.\\
\textbf{Step 6: Sampling} \\
Perform measurement operations on the targeted state.\\ 
Collect the samples of the prior and posterior probability distributions.\\ 
Perform the sampling using the pre-conditioned Crank-Nicholson $\rho_{CN}$ Metropolis-Hastings procedure or the Markov Chain Monte Carlo (MCMC) method.\\
\textbf{Step 7: Calculate Fidelity} \\
Compute the fidelity, $F$, between the predicted quantum state $|\psi_{\text{predicted}}\rangle$ and the target quantum state $|\psi_{\text{target}}\rangle$. \\
Compare the results of the prior and posterior distribution and analyse the results.\\
\textbf{Step 8: Output} \\
Plot the graph of the loss function, fidelities of the quantum circuit as shown in Fig~\ref{fig:bqst} (targeted and predicted state), and posterior probability distribution.
\end{algorithm}

\subsection{Quantum Variational Algorithm with Classical Statistics}
In this subsection, we discuss a variational algorithm suitable for quantum state reconstruction when measurement statistics of multiple Pauli strings are available. We call this algorithm the Quantum Variational Algorithm with Classical Statistics (QVCS).

Consider a pure state $\ket{\Phi}$ defined in terms of a certain basis consisting of vectors $\ket{\mathbf{n}_i}$. Without loss of generality, we can regard this as the computational basis, resulting in the following pure state representation
\begin{equation}
\ket{\Phi}=\sum_{i=0}^{N-1} C^i \ket{\mathbf{n}_i},
\end{equation}
where $N$ is the dimension of the Hilbert space. The complex coefficients, $C^i$, can be written in the polar form
\begin{equation}
C^i_{\lambda, \mu} = \braket{\mathbf{n}_i}{\Phi} \equiv \sqrt{p^i_\lambda} \exp\left(\imath\phi^i_\mu\right).
\label{quantum state}
\end{equation}
The parameters $\lambda$ and $\mu$ are introduced to indicate that the probability amplitudes and the phases are obtained in this algorithm through two different parameterised circuits. Following \cite{ref21}, in the classical approach, the parameters are obtained from the output nodes of two NNs.
\begin{figure}[h!]
\centering
\includegraphics[width=\linewidth]{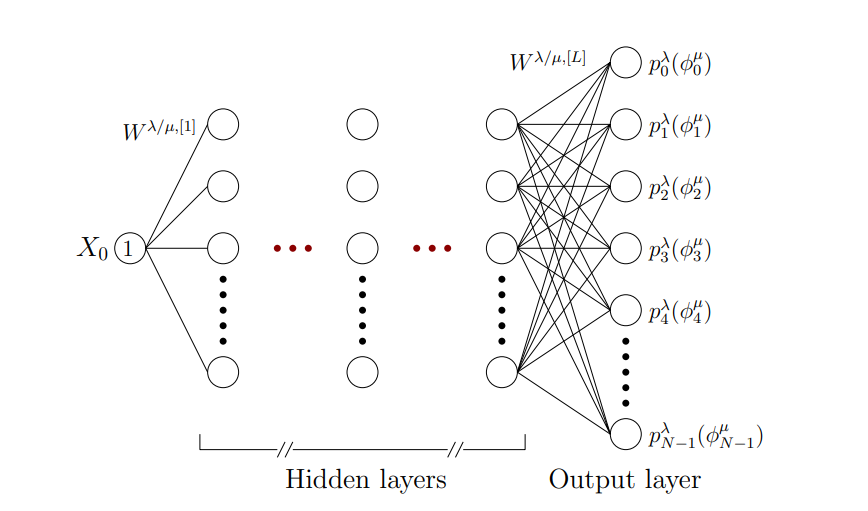}
\vspace{-0.8cm}
\caption{Perceptron model architecture to represent $p_\lambda$, and $\phi_\mu$ for the pure state $\ket{\phi}$, according to the parameterisation defined in Eq.\eqref{quantum state}. Here, the two networks, one for the probabilities and one for the phases, have been represented in a single diagram.}
\label{fig:neural network}
\end{figure}
Fig.\ref{fig:neural network} schematically shows how these parameters are obtained as the output nodes of two NNs. 

The $\lambda$ network outputs a probability vector $\{ p_\lambda^i \}_{i=0}^{N-1}$, while the $\mu$ network outputs phase vector $\{ \phi_\mu^i \}_{i=0}^{N-1}$. Since each network uses hidden layers, we have that
\begin{equation}
X_{\lambda / \mu}^{[l]} = \operatorname{ReLU}\pars{W_{\lambda / \mu}^{[l]} X_{\lambda}^{[l-1]} + b_{\lambda}^{[l]}}
\end{equation}
for $l = [1, L-1]$ and $X_{\lambda / \mu}^{[0]} = 1$. We then find that the $\lambda$ and $\mu$ networks output
\begin{equation}
\begin{aligned}
\cbrks{p_\lambda^i} &= \operatorname{softmax} \pars{W_{\lambda}^{[L]} X_{\lambda}^{[L-1]} + b_{\lambda}^{[L]}}, \\ 
\cbrks{\phi_\mu^i} &= \pi \tanh \pars{W_{\mu}^{[L]} X_{\mu}^{[L-1]} + b_{\mu}^{[L]}}, 
\end{aligned}
\end{equation}
respectively. Here, the $\operatorname{softmax}$ activation function has been used for the $\lambda$ network, while the $\mu$ network uses $\pi\tanh$. 

We propose an analogous algorithm where the hidden layers of the classical network are replaced by a Parameterised Quantum Circuit (PQC)/Quantum Neural Network (QNN). We use two kinds of circuit ansatz for the parameterised circuit. We call the circuits: circuit (A) and circuit (B) henceforth. Circuit (A) has been proposed in \cite{ref15} and was found to be expressive enough to encode the entanglement structure of quantum states to good fidelities using reasonable circuit depths of $10-17$ in their algorithm. Circuit (B) was the \verb|TwoLocal| form available in \verb|qiskit.circuits| library.

\begin{figure}[h]
\centering
\includegraphics[width=\linewidth]{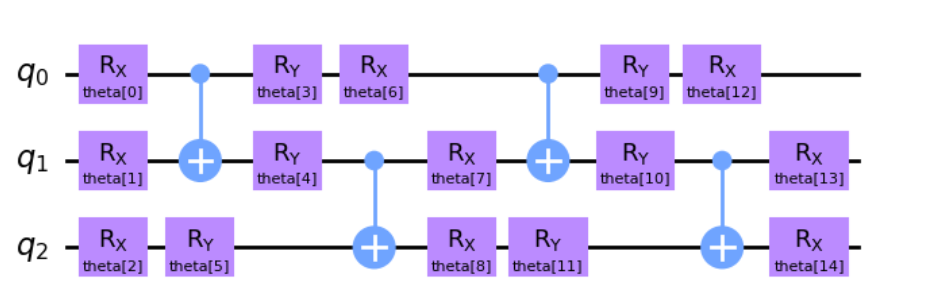}
\vspace{-0.6cm}
\caption{Circuit (A) for three qubits and depth five.}
\label{fig:circuit A}
\end{figure}

\begin{figure}[!h]
\centering
\includegraphics[width=\linewidth]{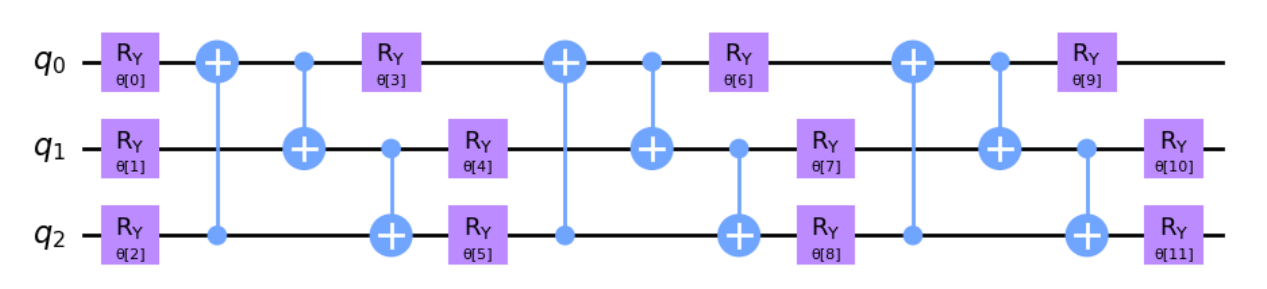}
\vspace{-0.6cm}
\caption{Circuit (B) for three qubits and depth four.}
\label{fig:circuit B}
\end{figure}

These are shown in Figs~\ref{fig:circuit A} and \ref{fig:circuit B} respectively. 

The number of output nodes in the NN for an $n$-qubit state is $2^n$. In the quantum circuit, we require $n$ qubits, and the outputs are obtained by measuring the probability distribution of all the $2^n$ output strings. Since sampling from this distribution will give only real numbers in the range $\left[0,1\right]$, to obtain the parameters $\phi^i$ from the circuit, we multiply by $2\pi$.

For the quantum circuit analogous to the NN in Fig~\ref{fig:neural network}, we assume that each quantum state is measured in a set of $N_B$ bases $\left\{ \ket{\mathbf{n}_0^{[b]}}, \ldots, \ket{\mathbf{n}_{N-1}^{[b]}} \right\}$, where $b \in \{ 0, \ldots, N_B - 1 \}$ is a basis index. We can then define the loss function $E$ used for this variational algorithm, which is of the maximum likelihood form
\begin{equation}
E \equiv-\sum_b \mathcal{L}_b=-\sum_b f^{[b]}_{\mathbf{n}_i} \log\left[\mathbb{P}(\mathbf{n}^{[b]}_i)\right],
\end{equation}
where $f^{[b]}_{\mathbf{n}_i}$ is the frequency of outcomes labeled by index $i$ for measurement in basis $b$, $\mathbb{P}(\mathbf{n}^{[b]}_i)$ refers to the probability of getting the output eigenvector $\mathbf{n}^{[b]}_i$ when measuring in the $b$ basis for a given state. Once the parameters of a state are given, this is simply calculated by the Born rule
\begin{equation}
\mathbb{P}(\mathbf{n}^{[b]}_i)=\abs{\bra{\mathbf{n}^{[b]}_i}\ket{\Phi}}^2. 
\end{equation}
Given that \begin{equation}
\abs{\bra{\mathbf{n}^{[b]}_i}\ket{\Phi}}^2 = \sum_{j=0}^{N-1} \braket{\mathbf{n}_i^{[b]}}{\mathbf{n}_j} C^j_{\lambda, \mu},
\end{equation}
we find that the loss function becomes
\begin{equation}
E = \sum_{b=0}^{N_B-1}\sum_{i=0}^{N-1}f^{[b]}_{\mathbf{n}_i} \log \left(\sum_{j=0}^{N-1}\bra{\mathbf{n}^{[b]}_i}\ket{\mathbf{n}_j}\sqrt{p_\lambda^j}e^{i\phi^j_\mu}+ \mathrm{c.c} \right).
\label{loss equation}
\end{equation}
The frequencies $f_{\mathbf{n}_i}$ were generated for our models using a \texttt{qiskit} backend \verb|qasm_simulator|, The measurement results for all Pauli strings were measured for a multi-qubit system, and the fidelities were benchmarked for random quantum states (see results section).

The algorithm as explained in  \ref{algo:qstqvcs_algorithm} can be summarised with the following steps:
\begin{itemize}

\item Firstly, a suitable dataset {$\mathcal{D}$} containing the measurement statistics of several Pauli strings needs to be obtained. To analyse our algorithm, we generate $\mathcal{D}$ via simulation through a \texttt{qiskit} backend.

\item The parameters ${\Theta_n}$ of the chosen variational circuit are initialised to random values. The number of parameters is $2\times\;\text{Depth}\;\times N_{\text{qubits}}$.

\item The circuits are sampled using \verb|SamplerQNN| class in \verb|qiskit_machine_learning.neural_networks| for getting $2^{N_{qubits}}$ output frequencies which are interpreted as the parameters ${p_\lambda^i}$ and ${\phi_\mu^i}$ respectively.

\item A classical optimiser, in this case, COBYLA, minimises the loss function defined according to Eq.\eqref{loss equation} above.

\item To calculate the gradients with respect to these parameters, we use the ``parameter shift rule'', as described in \cite{ref20}.

\item The final state parameters are obtained  after the optimisation process such that the observed frequencies $f^{[b]}_{\mathbf{n}_i}$ have a maximum likelihood.
\end{itemize}

Our analysis found that our proposed algorithm produced fidelities of roughly $70-80\%$ based on the circuit depth. However, the variance in the fidelities was also significant, pointing to the fact that the circuit ansatz was not complex enough to capture the entanglement structure of the state and the resulting fidelity when the states were highly entangled. To test this hypothesis, only states where one qubit was disentangled from the rest were given to the algorithm, and the fidelities achieved, along with the variance, were improved. 
 
\begin{algorithm}[t]
\caption{QVCS for QST}
\label{algo:qstqvcs_algorithm}

\textbf{Input:} Number of qubits, $N_{\text{qubits}}$, Circuit depth, $D_{\text{circ}}$.

\textbf{Step 1: Initialise Parameters} \\
Define the ansatz for the PQC as $\mathcal{C}$\; \\
Generate a random target quantum state $|\psi_{\text{target}}\rangle$.

\textbf{Step 2: Data Preparation} \\
Obtain measurement data $\mathcal{D}$ through a simulated measurement function, and normalise the measurement data $\mathcal{D}$ obtained from the target state.

\textbf{Step 3: Parameter Initialisation} \\
Initialise the parameter vector $\boldsymbol{\theta}$ , i.e.  ($2\times N_{\text{qubits}}\times D_{\text{circ}}$) with random values.

\textbf{Step 4: Optimisation} \\
Optimise the parameter vector $\boldsymbol{\theta}$ to minimise the discrepancy between the predicted state and the target state through the log-likelihood function. 

\textbf{Step 5: Extract Optimised Parameters} \\
Reshape the optimised parameter vector $\boldsymbol{\theta}$ into a matrix $\boldsymbol{\Theta}$.

\textbf{Step 6: Predict Quantum State} \\
Calculate the predicted quantum state amplitudes $|\psi_{\text{predicted}}\rangle$ using $\mathcal{C}$ and the optimised parameters $\boldsymbol{\Theta}$.

\textbf{Step 7: Calculate Fidelity} \\
Compute the fidelity $F$ between the predicted quantum state $|\psi_{\text{predicted}}\rangle$ and the target quantum state $|\psi_{\text{target}}\rangle$.

\textbf{Step 8: Output} \\
Present the optimised PQC, the predicted quantum state amplitudes $|\psi_{\text{predicted}}\rangle$, and the fidelity as the outcomes of the algorithm.
\end{algorithm}
\section{\label{sec:level3}Results and Discussion} \hypertarget{sec5}{}
In this section, we present the outcomes of our research experiments conducted using the \texttt{AerSimulator} \cite{ref91}. We detail the results of utilising a pre-built quantum model made available through the \texttt{qiskit} framework \cite{ref90}. Subsequently, we delve into the outcomes derived from applying four distinct quantum methods, as previously expounded: The VQC algorithm, the qPCA algorithm, the Bayesian approach, and the QVCS algorithm.

\begin{figure}[h!]
\centering
\includegraphics[width=\linewidth]{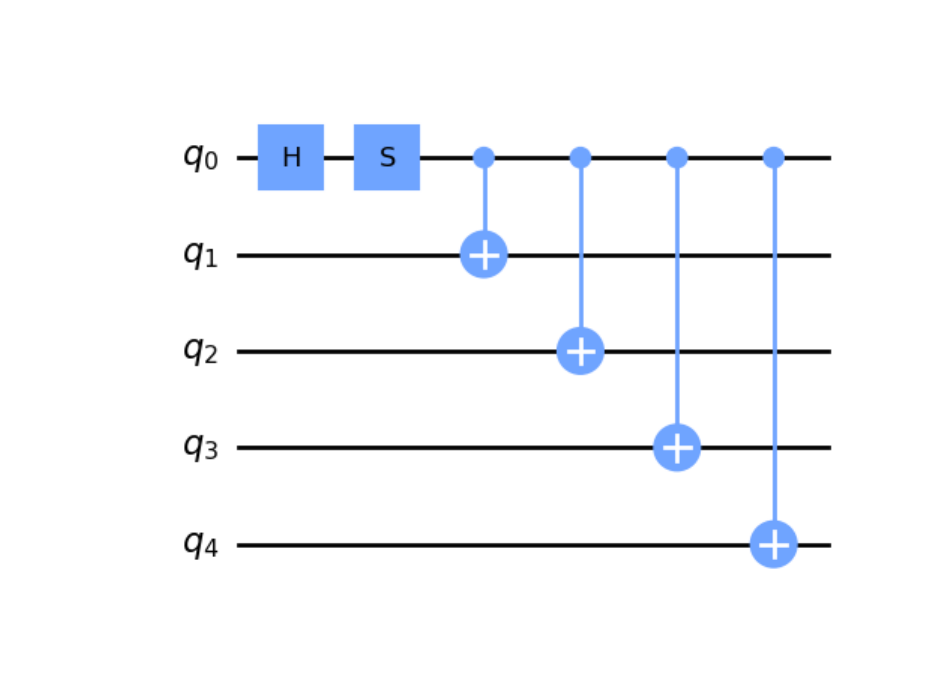}
\vspace{-1.5cm}
\caption{Circuit diagram for GHZ state generation.}
\label{fig:GHZ}
\end{figure}

Starting with the pre-built model, we used the \texttt{StateTomography} Experiment from \texttt{qiskit} to construct a circuit to generate the Greenberger-Horne-Zeilinger (GHZ) state \cite{ref101}. The circuit utilised Hadamard and controlled-X gates to create entanglement among the qubits, as shown in Fig~\ref{fig:GHZ}.

Subsequently, we performed a \texttt{StateTomography} experiment on the circuit. This experiment involved measuring the qubit's states in various bases, which allowed us to reconstruct the density matrix of the GHZ state. The average state fidelity was $98.38 \%$, indicating that the circuit could generate the GHZ state with high fidelity.
 
In addition, we performed a parallel tomography experiment on the circuit, measuring the state of the qubits in various bases, but it was only performed on individual circuit gates, which allowed us to investigate the discrete influence of each gate on the state of the qubits.

\begin{figure}[h!]
\centering
\includegraphics[width=\linewidth]{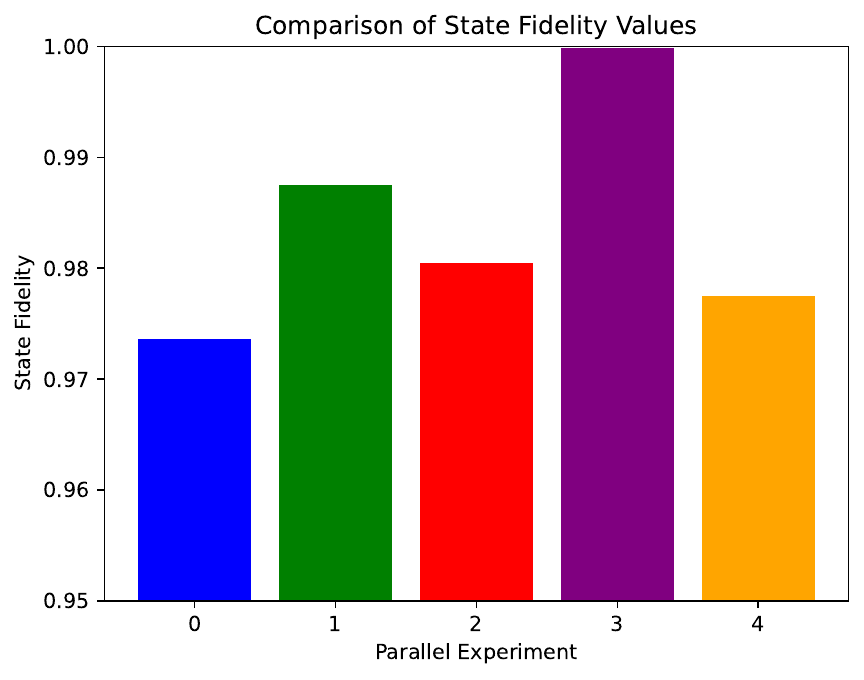}
\vspace{-0.6cm}
\caption{State fidelity comparison plot for the QST experiment - qiskit.}
\label{fig:state_fidelity}
\end{figure}

The results of the parallel tomography experiment showed that all of the gates in the circuit had a high fidelity, as shown in Fig~\ref{fig:state_fidelity}, which indicates that the circuit could generate the GHZ state with high fidelity, even when individual gates were not perfect.

\begin{figure}[h!]
\centering
\includegraphics[width=\linewidth]{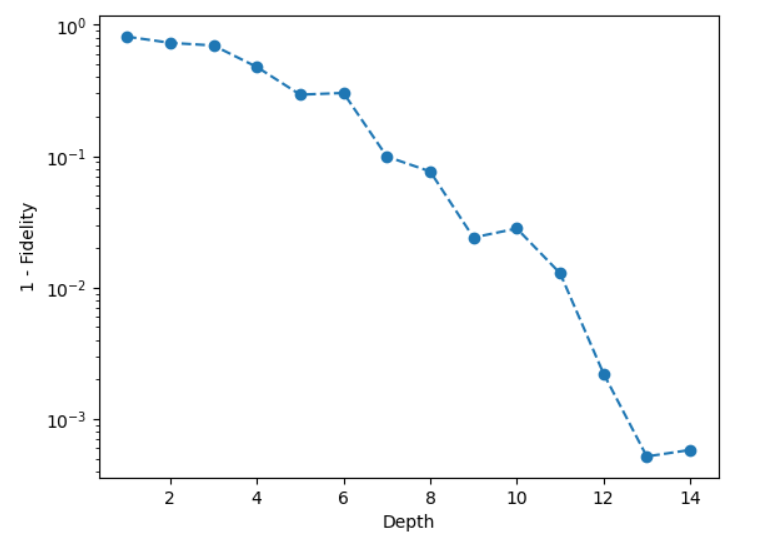}
\vspace{-1cm}
\caption{Correlation between loss Values and circuit depth in the VQC algorithm.}
\label{fig:vqcdepth}
\end{figure}

\begin{figure}[h!]
\centering
\includegraphics[width=\linewidth]{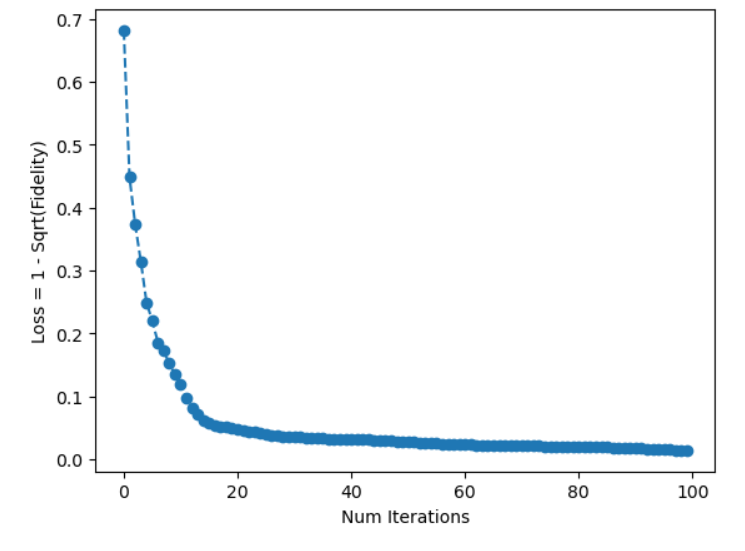}
\vspace{-1cm}
\caption{Correlation between loss values and iterations in the VQC algorithm.}
\label{fig:vqciteration}
\end{figure}

Secondly, the VQC algorithm was effectively applied to mixed states, as discussed in the previous section. Our investigation explored the intricate relationship between the loss value and two distinct parameters: Depth and iteration numbers, as shown in Figs~\ref{fig:vqcdepth} and \ref{fig:vqciteration}, respectively. The results of our study clearly show that an optimal fidelity value was obtained at a depth of $13$. Additionally, a compelling trend emerges: Increasing the depth number correlates with a decrease in the loss value, effectively approximating the ground states of a local spin Hamiltonian.


Notably, the results highlight the algorithm's ability to achieve high fidelity using a relatively small number of variational parameters and iterations. This observation attests to the efficacy of the VQC approach in realising accurate quantum computations. Furthermore, the proposed algorithm's potential extends to near-term quantum devices, presenting a promising avenue for future research exploration.

In our next method, the qPCA, we investigated how the accuracy of the reconstructed state $\rho_\text{reconstructed}$ varies with $t$ by initialising the algorithm with a known random state $\rho_\text{original}$ and calculating the average fidelity from $N_\text{iter}=50$ iterations of QPE for a $t$ value of $t = t_i$. 

\begin{figure}[h!]
\centering
\includegraphics[width=\linewidth]{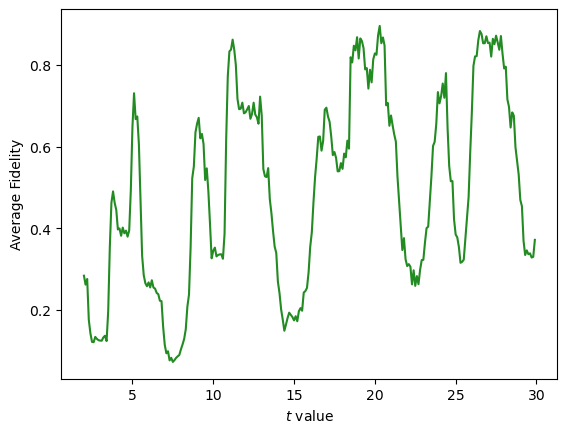}
\vspace{-0.8cm}
\caption{Average fidelity $\bar{F_i}$ between States $\rho_\text{original}$ and $\rho_{\text{reconstructed}}$ for a randomised $\rho_\text{original}$ and a 2-qubit phase estimation with $t = t_i$.} 
\label{fig:qPCA_avg}
\end{figure}

\begin{figure}[h!]
\centering
\includegraphics[width=\linewidth]{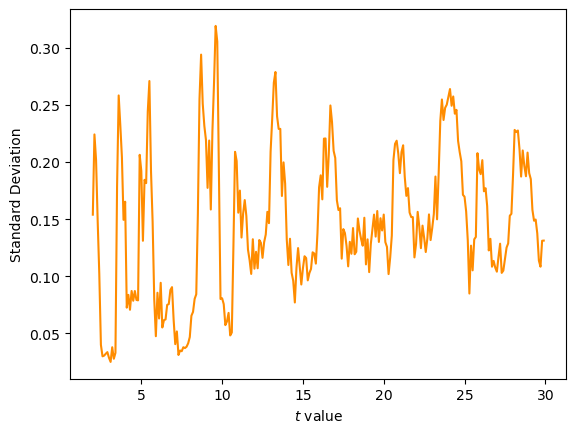}
\vspace{-0.8cm}
\caption{Standard Deviation $\sigma_i$ of the calculated Fidelities over $50$ iterations between States $\rho_\text{original}$ and $\rho_{\text{reconstructed}}$ for a randomised $\rho_\text{original}$ and a 2-qubit phase estimation with $t = t_i$.} 
\label{fig:qPCA_std}
\end{figure}

Each iteration yields a different fidelity $F_j$ for the $j^{\text{th}}$ reconstructed state. The resulting values are then used to compute the average fidelity $\bar{F_i}$ and the standard deviation $\sigma_i$ for each $t_i$. These average fidelity and standard deviation values are depicted in Figs~\ref{fig:qPCA_avg} and \ref{fig:qPCA_std}, respectively. Here, QPE was performed using $2$ auxiliary qubits and calculated for $t_i \in T = \cbrks{t \ | \ t = 2 + 0.1n, \ 0 \leq n < 280}$. 

Using the results of our calculations, we can then find the optimal $t$ value and its associated average fidelity, which is the maximum average fidelity. Let $\mathcal{F}(t)$ be a function that maps a value $t$ to its corresponding average fidelity. Then, the maximum average fidelity $\bar{F}_\text{max} \approx \arg \max _{t \in T} \mathcal{F}(t)$. We find that the optimal $t$ value is $t_\text{optimal} \approx 20.3$ with an associated average fidelity of $\bar{F}_\text{max} \approx 89.58 \%$ and a standard deviation of $\sigma_{t_\text{optimal}} = 0.10$. Using these statistics, we can establish a $95 \%$ confidence interval of ($86.76 \%$, $92.40 \%$) for the optimal $t$ value's fidelity.

\begin{figure}[!t]
\centering
\includegraphics[scale=0.5]{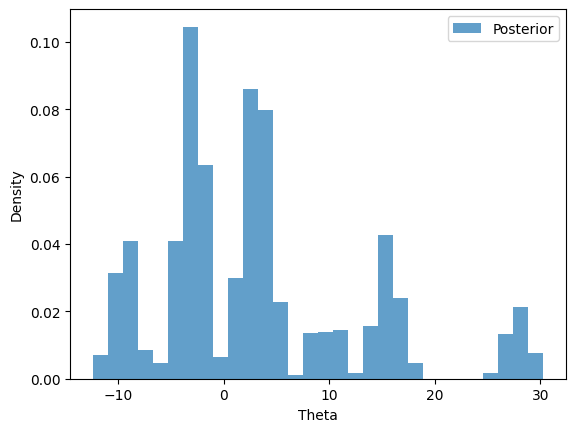}
\vspace{-0.2cm}
\caption{Posterior probability distribution for Bernoulli model in Bayesian approach for $\textit{N}$ = $4$ \textit{qubits} with a circuit depth = $8$. Here $\theta$ = $\theta'$ for posterior parameters on x-axis and $\rho(\theta)$ on y-axis.}
\label{fig:bayesian}
\end{figure}

Our next method used was the Bayesian approach which employs the Bernoulli model, where the posterior distribution is based on the evidence and likelihood functions. The likelihood $\mathbb{P}(\mathcal{D}|\boldsymbol{\theta})$ and evidence $\mathcal{P}(D)$ are taken as input, and the behaviour of prior distribution $\mathcal{P}(D)$ and the evidence $\mathcal{D}$ will affect the posterior probability. To optimise the set of parameters $\boldsymbol{\theta}'$, we used the COBYLA optimiser with a \verb|maxiter=150|. Fig~\ref{fig:bayesian} shows the results of our analysis for a system of $N = 4$ qubits and circuit depth range from $3$ to $8$. We ran $10\,000$ samples with $200$ learning steps for each sample. The parameters are updated after the sampling process is complete, and these posterior evidence then act as prior parameters to create a new set of posterior beliefs. The Markov Chain Monte Carlo (MCMC) Metropolis-Hastings procedure was used to sample on $4$-qubit data. We computed the fidelities after the sampling process.

\begin{figure}[!t]
\centering
\includegraphics[scale=0.5]{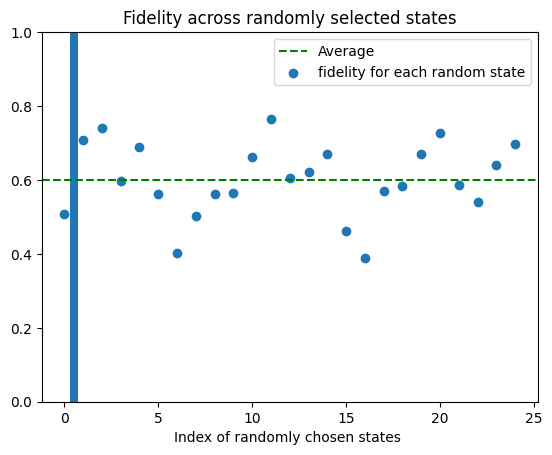}
\vspace{-0.2cm}
\caption{Fidelity Graph for Bayesian Inference with $N_{\text{qubits}}=4$. Each Fidelity is represented by blue dots, and benchmark fidelity with the green line. The results are plotted for circuit depth of range $8$ for $n = 25$ randomly generated states. } 
\label{fig:bayesianfidelity}
\end{figure}

Fig. \ref{fig:bayesianfidelity} shows the algorithm's performance on a simulator with $8$GB of RAM. The average fidelity was calculated for $n = 25$ randomly generated states, with circuit depth range $3-8$ and, and $n = 4$ qubits. For each circuit depth, a total of $25$ states are randomly generated, each state has its own fidelity. The plot shows the fidelity with blue dots versus the index of randomly chosen states for a circuit of $\text{depth}=8$. The performance can be improved by optimising the parameters and decreasing the circuit depth.


\begin{figure}[h!]
\centering
\includegraphics[width=\linewidth]{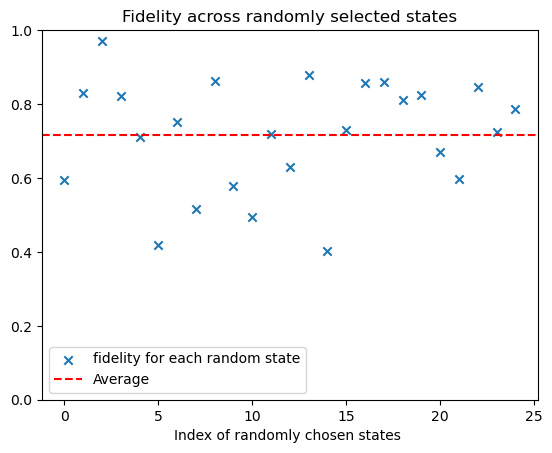}
\vspace{-0.8cm}
\caption{Fidelities for circuit (A) with $N_{\text{qubits}}=3$ and $D_{\text{circ}}=3$.}
\label{fig:benchmark}
\end{figure}

For our last algorithm, the Quantum Variational Algorithm with Classical Statistics, we evaluated its accuracy by calculating the average fidelity across $25$ randomly generated quantum states for a fixed circuit depth $D_{\text{circ}}$ and the number of qubits $N_{\text{qubits}}$. We used the classical optimiser COBYLA with the parameters \verb|rhobeg=30| and \verb|maxiter=150|.
The plot shown in Fig~\ref{fig:benchmark} demonstrates, for instance, how the algorithm performs on average for $D_{\text{\text{circ}}}=3$ and $N_{\text{\text{qubits}}}=3$. The circuit used here was circuit (A).

\begin{figure}[h!]
\centering
\includegraphics[width=\linewidth]{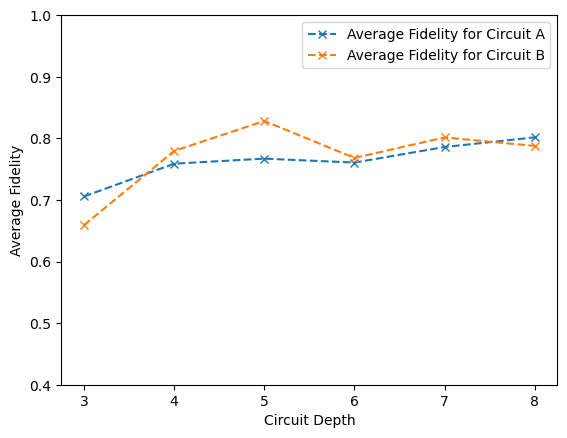}
\vspace{-0.8cm}
\caption{Comparison of average fidelities for $N_{\text{qubits}}=3$ in circuits (A) and (B).}
\label{fig: 3 qubits}
\end{figure}

Fig.\ref{fig: 3 qubits} shows the variation of average fidelity with circuit depth for both circuit ansatz (A) and (B). The bar plot of the standard deviation of the fidelities is shown in Fig~\ref{fig:standarddeviation} the standard deviation of the fidelities as a function of $D_{\text{circuit}}$ for $N_{\text{qubits}}=3$.
 	
 \begin{figure}[h!]
 \centering
\includegraphics[width=\linewidth]{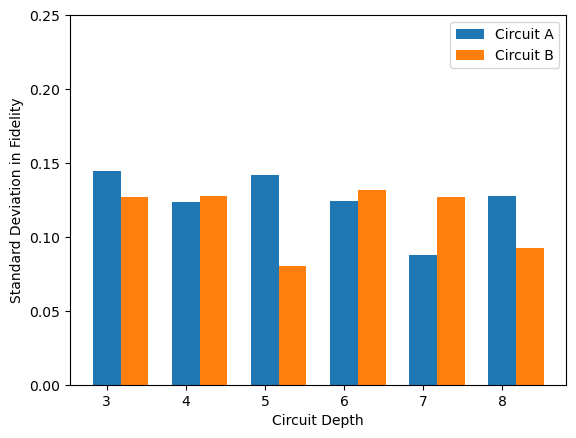}
 \vspace{-0.8cm}
\caption{Comparison of standard deviations in fidelities achieved by QVCS algorithm for $N_{\text{qubits}}=3$ in circuits (A) and (B).}
\label{fig:standard deviation}
\end{figure}

These results show that the average fidelity increases with circuit depth for both circuit ansatzes. However, the standard deviation of the fidelities exhibits a non-monotonic behaviour for circuit ansatz (B), decreasing at intermediate circuit depths before increasing again at deeper circuit depths. This behaviour is likely due to the interplay between the noise affecting the circuit ansatz (B) and the ability of the circuit ansatz to compensate for the noise.

\begin{figure}[h!]
\centering 
\includegraphics[width=\linewidth]{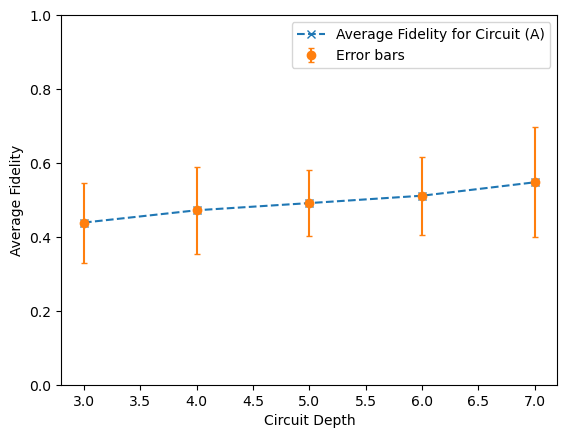}
\vspace{-0.8cm}
\caption{Fidelities and error bars for $N_{\text{qubits}}=4$ with parameterised circuit (A).}
\label{fig:4 qubit}
\end{figure}

The results for the circuit (A) are shown in Fig~\ref{fig:4 qubit}. The average fidelity for the circuit (A) with $4$ qubits is lower than that for $3$ qubits. This is likely because the noise affecting the circuit ansatz (A) is more pronounced with $4$ qubits.


\section{\label{sec:level4}Conclusion} \hypertarget{sec6}{}
In conclusion, this paper has presented a review of the various classical and quantum approaches to QST. The goal was to explore the various QML techniques and to show the computational advantage and improved accuracy over the classical approaches.

Our approach builds on the existing literature on QST and QML. In addition, we have provided a comprehensive literature review of some important works in the field. In sufficient detail, we have described the methods employed and the novelties of each approach. The classical approaches to state tomography were discussed by giving detailed descriptions of the theoretical underpinnings of each method. Furthermore, we have discussed the mathematical, computational, and circuit architecture of the QST methods and presented the results of each method employed by implementing them and verifying their fidelity.

It is important to point out the disadvantages of the QML approach. Firstly, in this current NISQ era of quantum devices, we only have access to a few qubits and thus are restricted to reconstructions of small systems. Secondly, the methods are not infallible against noise and decoherence, and thus, can lead to many potentially erroneous results. Thirdly, implementing these methods requires domain expertise which might be challenging for the non-expert to implement and validate their findings.

For our future work, we plan to explore several avenues of research to improve our QML-based QST methods further. Firstly, we will investigate the use of more advanced QML algorithms and consider quantum versions of the state-of-the-art classical methods to improve the accuracy and efficiency of our approach. Secondly, we will explore the use of hybrid classical-quantum algorithms to reduce further the number of measurements required for QST.

Overall, QML-based QST methods represent a significant advancement in the field of QIP and can potentially revolutionise how we perform state tomography. By leveraging the power of QML algorithms, we have shown that it is possible to achieve high accuracy with significantly fewer measurements than conventional methods. This has important implications for practical QIP applications, where the ability to perform QST efficiently and accurately is critical.


\end{document}